\documentclass[iop]{emulateapj}

\usepackage{color}
\usepackage[colorlinks,linkcolor=black,anchorcolor=black,citecolor=blue,hyperfootnotes=true]{hyperref}
\usepackage[section]{placeins}
\begin{document}

\title{A Model-independent Measurement of the Spatial Curvature using Cosmic Chronometers and the HII Hubble Diagram}
\author{Jing Zheng\altaffilmark{1}, Fulvio Melia\altaffilmark{2}, Tong-Jie Zhang\altaffilmark{1}}
\affil{$^1$Department of Astronomy, Beijing Normal University, Beijing 100875,China; 201511160108@mail.bnu.edu.cn,
tjzhang@bnu.edu.cn;}
\affil{$^2$Department of Physics, The Applied Math Program, and Steward Observatory, The University of Arizona, Tucson, AZ 85721, USA; fmelia@email.arizona.edu}

\begin{abstract}
We propose a model-independent way to determine the cosmic curvature using the Hubble parameter $H(z)$ 
measured with cosmic chronometers and the comoving distance $D(z)$ inferred from HII galaxies. We 
employ Gaussian processes to smooth the measure of distance and match it to $30$ values of $H(z)$. 
The curvature parameter $\Omega_k$ may be obtained individually for each such pair. The weighted average 
for the complete sample is $\Omega_k=-0.0013\pm0.0004$, suggesting a bias towards negative
values. The accuracy of the curvature measurement improves with increased redshift, however, given
possible systematic effects associated with local inhomogeneities. We therefore also analyze a
high-redshift ($z>1.5$) sub-sample on its own, which is more likely to reflect the geometry of the 
Universe on large, smooth scales. We find for this set of data that $\Omega_k=-0.0111\pm0.0416$,
consistent with zero to better than $1\sigma$. This result is in agreement with the spatially flat 
universe inferred from the cosmic microwave background observations. We expect this method to
yield even tighter constraints on the curvature parameter with future, more accurate observations
of HII galaxies at high $z$.
\end{abstract}

\keywords{cosmological parameters --- cosmology: observations --- HII regions: galaxies:general}

\section{Introduction}
The spatial curvature constant of the universe is one of the most fundamental and vital parameters 
in modern cosmology. Estimating whether the universe is open, flat, or closed is a robust way to test 
the homogeneous and isotropic Friedmann-Lema\^itre-Robertson-Walker (FLRW) metric. It is also closely 
related to many important problems, such as the evolution of the Universe and the nature of dark energy. 
A significant detection of nonzero curvature would produce far-reaching consequences for fundamental 
physics and inflation theory\citep{ich1,cla1,gw,vir1}. As of today, however, a flat Universe is supported 
by most of the observational data, including the latest Planck result \citep{plank1}. 

There exists a strong degeneracy between cosmic curvature and the dark-energy equation of state,
however, so we cannot directly constrain these two quantities simultaneously. In general, the universe 
is assumed to be flat for the analysis of dark energy, or dark energy is assumed to be a
cosmological constant for the determination of curvature. But such simplified assumptions may 
result in errors and confusion, even though the true curvature is known to be very small 
\citep{cla1,vir1}. In \citet{cla1,cla2}, when arguing the defects of a zero curvature assumption, 
a direct curvature determination method was proposed via combining measurements to yield the
expression
\begin{equation}
\Omega_k=\frac{[H(z){D}'(z)]^{2}-c^{2}}{[H_0D(z)]^{2}}\;,\label{equ:1} 
\end{equation}
where $D^\prime(z)\equiv dD(z)/dz$, $c$ is the speed of light, and $H_0$ is the Hubble constant. 
The quantity $D(z)$ is the comoving distance, which may be expressed alternatively in terms
of the angular-diameter distance, $D(z)=(1+z)D_A(z)$, or the luminosity distance,
$D(z)=D_L(z)/(1+z)$ \citep{hog}. Using this equation, one may measure the cosmic curvature 
with a pair of observations yielding the redshift-dependent distance and rate of expansion
in a model independent way, without the need of introducing other model parameters or a
dark-energy model. In principle, one may use this approach to measure the curvature at any
single redshift though, of course, the accuracy improves with a statistical weighting over
a large redshift range.

Since its proposal, this method has been used on many occasions \citep{sc,mj,li1,sap1,yah1,cai1,ran1}. 
For example, one may extract the necessary data from Baryon Acoustic Oscillations (BAO) \citep{yw}, 
or Type Ia SNe \citet{suz} and strong gravitational lenses from the Sloan Lens ACS Survey \citet{bol,lzx,zzh}. 
In this paper, we propose a new method based on the measurement of distances with HII galaxies,
and the expansion rate with cosmic chronometers. We first reconstruct the distance modulus as
a function of redshift $z$ using Gaussian Process, from which we can calculate a continuous comoving 
distance and its error. This function may be evaluated at the $30$ individual redshifts where $H(z)$ 
has been measured. In principle, the cosmic curvature may therefore be measured at $30$ different
redshifts. 

The structure of this paper is as follows. The data are described in \S~2, which includes the 
most recent observations of HII galaxies and cosmic chronometers. In \S~3, we introduce the 
methodology of our model-independent approach for constraining $\Omega_k$. The results and 
discussion are presented in \S~4, and we conclude \S~5.

\section{Observational Data}
\subsection{Hubble parameter from cosmic chronometers}
The Hubble parameter $H(z)$ provides the redshift-dependent expansion rate of the universe.
In recent years, $H(z)$ has been measured using two approaches: (1) via the detection of 
radial BAO features \citep{gaz1,bla1,sam1} which, however, requires the adoption of a 
particular cosmological model, e.g., to disentangle the BAO signal from contamination
due to internal redshift space distortions; (2) the so-called cosmic-chronometer approach,
in which one calculates the value of $dz/dt$ (the derivative of redshift relative to
cosmic time), as first proposed by \citet{jl2}. This quantity can be obtained 
by measuring the differential age of two red galaxies at different redshifts. The
distinct advantage of method (2) over (1) is that no cosmology needs to be assumed, 
providing a model-independent evaluation of $H(z)$:
\begin{equation}
H(z)=-\frac{1}{(1+z)}\frac{dz}{dt}\approx-\frac{1}{(1+z)}\frac{\Delta z}{\Delta t}\;.\label{equ:2} 
\end{equation}
Since the cosmic-chronometer approach is the only model-independent method of determining 
$H(z)$, we shall use it exclusively for the analysis in this work. The $30$ measurements
of $H(z)$ from \citet{lm} are listed in Table~\ref{table:1}. These are chosen from the
compilation in \citet{zheng1}, though omitting those values obtained with BAO. 

\subsection{HII galaxies}
The HII galaxy sample used in this paper comprises 156 sources in total, including 25 high-$z$ HII galaxies, 
107 local HII galaxies, and 24 giant extragalactic HII regions \citep{ter1}. For these sources, one
determines the line fluxs $F(H\beta)$ and gas velocity dispersion $\sigma(H\beta)$, which may be used together 
with the correlation $L(H\beta)-\sigma(H\beta)$ between the luminosity and $\sigma(H\beta)$ \citep{cha1,cha2,ter1}
to infer the luminosity distance:
\begin{equation}
\log L(H\beta)=\alpha \log\sigma(H\beta)+\kappa\;,\label{equ:3}
\end{equation} 
where $\sigma(H\beta)$ is the velocity dispersion of the $H\beta$ line, $\alpha$ is the slope and 
$\kappa$ is the (constant) intercept (i.e., the logarithmic luminosity at $\log \sigma(H\beta)= 0$). 
Although the coefficients $\alpha$ and $\kappa$ may be model-dependent, so that one would
normally need to optimize them simultaneously with the cosmological parameters, \citet{wei1} 
have shown that these parameters are very insensitive to the underlying cosmology. Therefore, 
we may directly use the average optimized values in \citet{wei1}, which are 
$\alpha=4.87^{+0.11}_{-0.08}$ and $\sigma = 32.42^{+0.42}_{-0.33}$. The variation of
these quantities between different cosmological models is smaller than the quoted error. 

\begin{table}
\begin{center}
\caption{30 Hubble Parameter measurements using cosmic chronometers}
\renewcommand\tabcolsep{0.5pt}{
\begin{tabular}{cccl}
\hline
Redshift & $H(Z)$ & $\sigma_{H}$ & Reference \\
 &($\mathrm{km\ s^{-1}\ Mpc^{-1}}$)&($\mathrm{km\ s^{-1}\ Mpc^{-1}}$)& \\
\hline
0.07 &69 &19.6 &\citet{zhang1}\\
0.09 &69 &12 &\citet{jim1}\\
0.12 &68.6 &26.2 &\citet{zhang1}\\
0.17 &83 &8 &\citet{sim1}\\
0.1791 &75 &5 &\citet{mor1}\\
0.1993 &75 &5 &\citet{mor1}\\
0.2 &72.9 &29.6 &\citet{zhang1}\\
0.27 &77 &14 &\citet{sim1}\\
0.28 &88.8 &36.6 &\citet{zhang1}\\
0.3519 &83 &14 &\citet{mor1}\\
0.3802 &83 &13.5 &\citet{mor1}\\
0.4 &95 &17 &\citet{sim1}\\
0.4004 &77 &10.2 &\citet{mor2}\\
0.4247 &87.1 &11.2 &\citet{mor2}\\
0.4497 &92.8 &12.9 &\citet{mor2}\\
0.4783 &80.9 &9 &\citet{mor2}\\
0.48 &97 &62 &\citet{ste1}\\
0.5929 &104 &13 &\citet{mor1}\\
0.6797 &92 &8 &\citet{mor1}\\
0.7812 &105 &12 &\citet{mor1}\\
0.8754 &125 &17 &\citet{mor1}\\
0.88 &90 &40 &\citet{ste1}\\
0.9 &117 &23 &\citet{sim1}\\
1.037 &154 &20 &\citet{mor1}\\
1.3 &168 &17 &\citet{sim1}\\
1.363 &160 &33.6 &\citet{mor3}\\
1.43 &177 &18 &\citet{sim1}\\
1.53 &140 &14 &\citet{sim1}\\
1.75 &202 &40 &\citet{sim1}\\
1.965 &186.5 &50.4 &\citet{mor3}\\
\hline
\end{tabular}}\label{table:1}
\end{center}
\end{table}

Then, using the $L(h\beta)-\sigma(H\beta)$ relation, the distance modulus of an HII galaxy 
is given as 
\begin{equation}
\mu_{\rm obs} = -\sigma(H\beta) + 2.5[\alpha \log\sigma(H\beta) - \log F(H\beta)\;.]\label{equ:4} 
\end{equation}
The error $\sigma_{\mu_{\rm obs}}$ in $\mu_{\rm obs}$ is calculated as
\begin{equation}
\sigma_{\mu_{\rm obs}} = 2.5[(\alpha \sigma_{\log\sigma[H\beta]})^{2}+(\sigma_{\log F[H\beta]})^{2}]^{1/2}\;.\label{equ:5} 
\end{equation}
For convenience, one may combine $\kappa$ and the Hubble constant $H_0$ together as \citet{wei1}
\begin{equation}
\sigma(H\beta) = -2.5\kappa -5\log_{10}H_{0} + 125.2\;.\label{equ:6}
\end{equation} 
Correspondingly, the theoretical distance modulus is given as 
\begin{equation}
\mu=5\log[\frac{\tilde{D}_L(z)}{\rm Mpc}]\;,\label{equ:7}
\end{equation} 
where $\tilde{D}_L(z) \equiv H_0D_L(z)$. In our calculations, we adopt a Hubble constant $H_0 = 70$ 
km s$^{-1}$ Mpc$^{-1}$, but the actual value of $H_0$ does not affect the final result because 
it cancels out from the various expressions. 

\section{Gaussian Processes (GP)}
\begin{figure}[hp]
\centering
\includegraphics[scale=0.60]{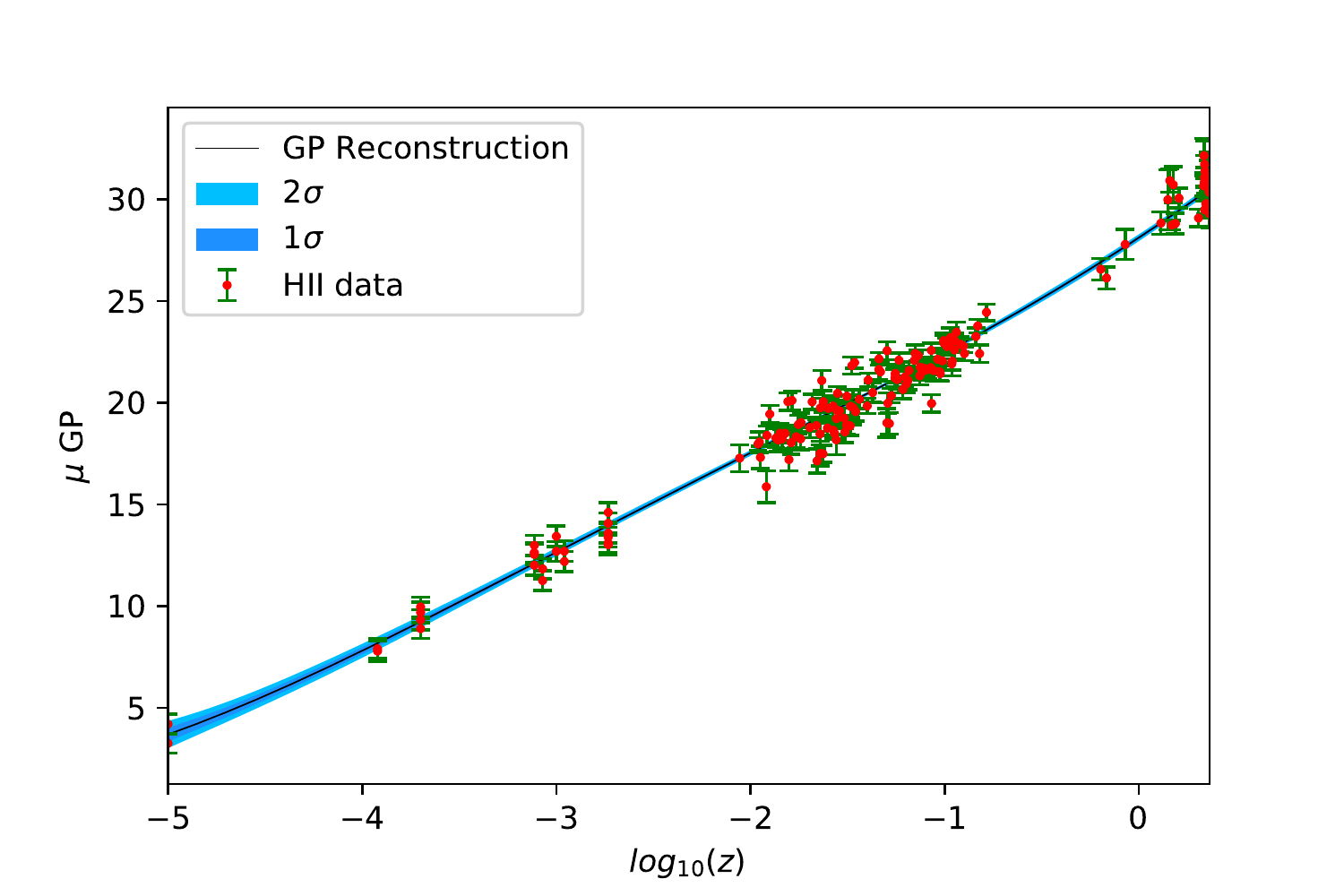}
\includegraphics[scale=0.60]{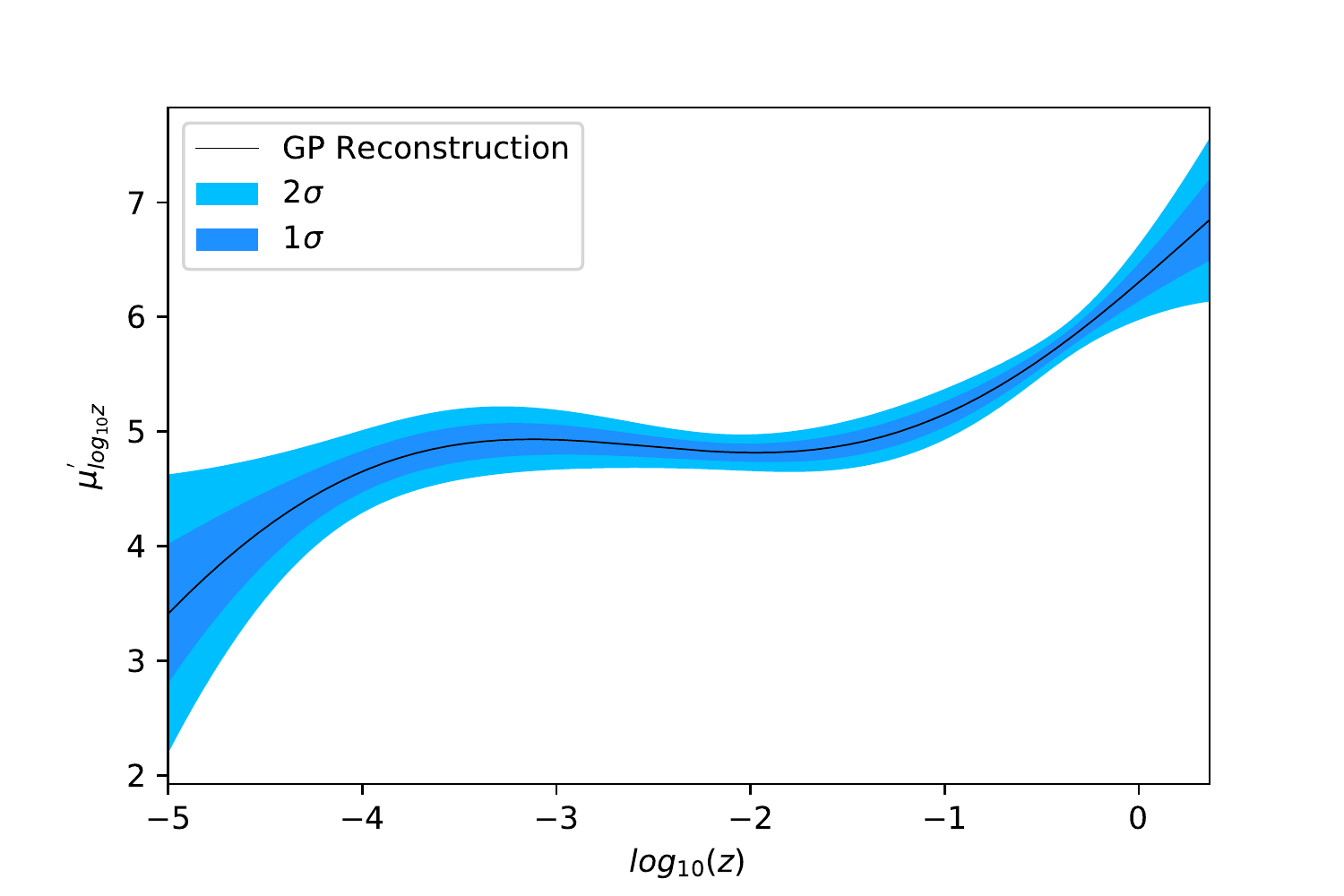}
\caption{Top: (Solid black) The GP reconstruction of the distance modulus $\mu_{GP}(\log_{10}[z])$, 
based on equation~(\ref{equ:4}) using HII galaxy data. The sample contains 156 HII-regions and Galaxies, 
shown as red circles with $1\sigma$ error bars. The dark blue swath represents the $1\sigma$ confidence 
region of the reconstruction, while light blue is $2\sigma$. Bottom: (Solid black) The GP reconstruction 
of the first derivative $\mu^{'}_{GP}(\log_{10}[z])$. The shaded swaths have the same meaning as in
the top panel.}\label{fig:GP}
\end{figure}

\begin{figure}[hp]
\centering
\includegraphics[scale=0.60]{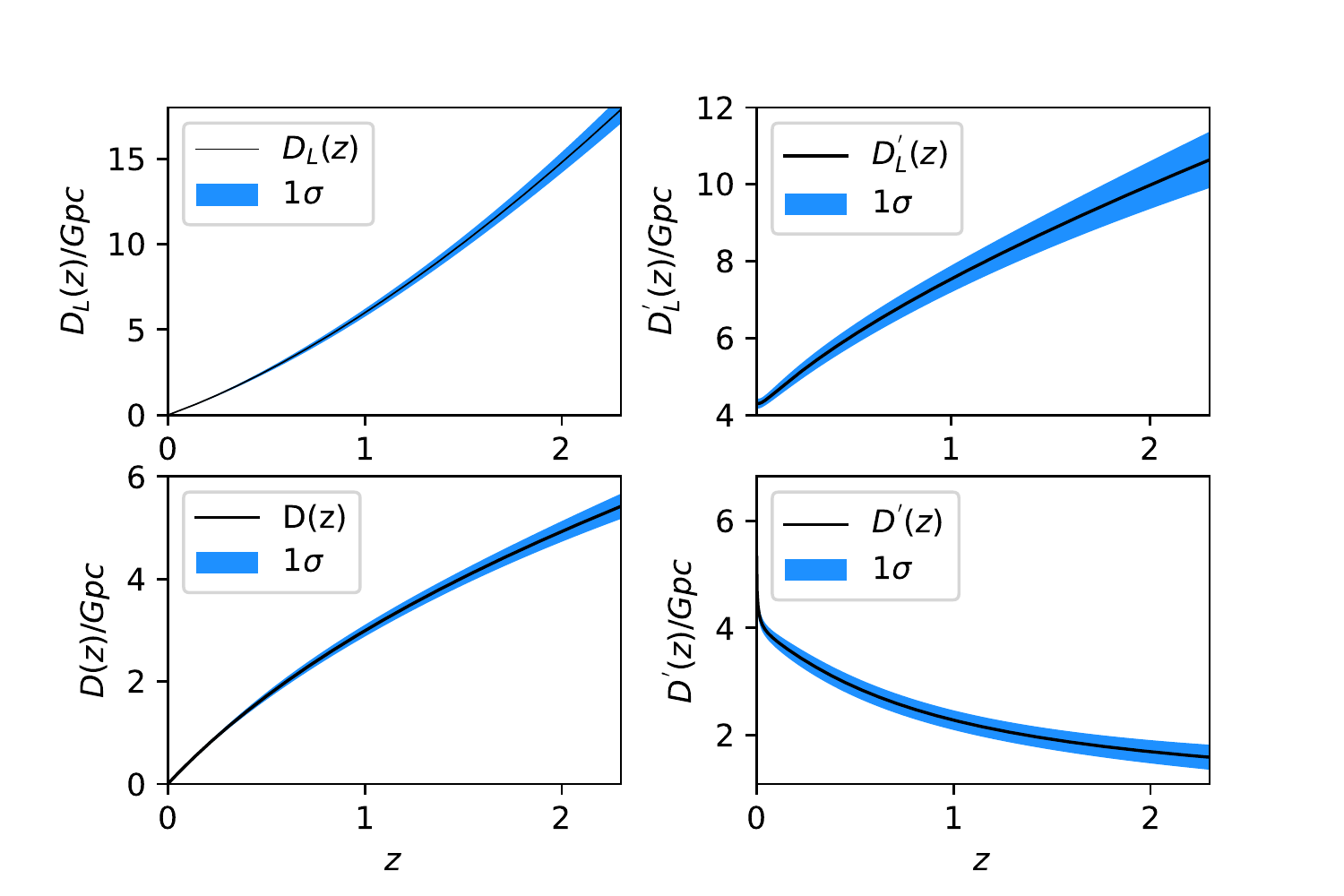}
\caption{(Solid black) The continuous luminosity distance (top-left) and its first derivative (top-right), and 
comoving distance (bottom-left) and its first derivative (bottom-right). The shaded swaths show the 
$1\sigma$ confidence regions.}\label{fig:4pic}
\end{figure}

To reconstruct the distance modulus $\mu_{GP}$ as a function of redshift from equation~(\ref{equ:4}), 
we use the GP code\footnote{http://www.acgc.uct.ac.za/~seikel/GAPP/index.html} developed by \citet{sei1}. 
This algorithm provides both the distance modulus and its error directly. The results of this
reconstruction are shown in figure~\ref{fig:GP}, which includes both the distance modulus 
$\mu_{GP}(\log_{10}[z])$ and its first derivative $\mu^{'}_{GP}(\log_{10}[z])$. In the last step,
we to convert $\log_{10}(z)$ into the redshift $z$ and thereby calculate $\mu$ from $\mu_{GP}$. 
From equation~(\ref{equ:7}), one then infers the luminosity distance, its first derivative and 
their errors, from the `measured' quantities $\mu$, $\mu^{'}$ and $\sigma_{\mu}$ and $\sigma_{\mu^{'}}$.
The outcome of this procedure is displayed in figure~\ref{fig:4pic}.

\section{Results and Discussion}
\begin{figure}[h]
\centering
\includegraphics[scale=0.49]{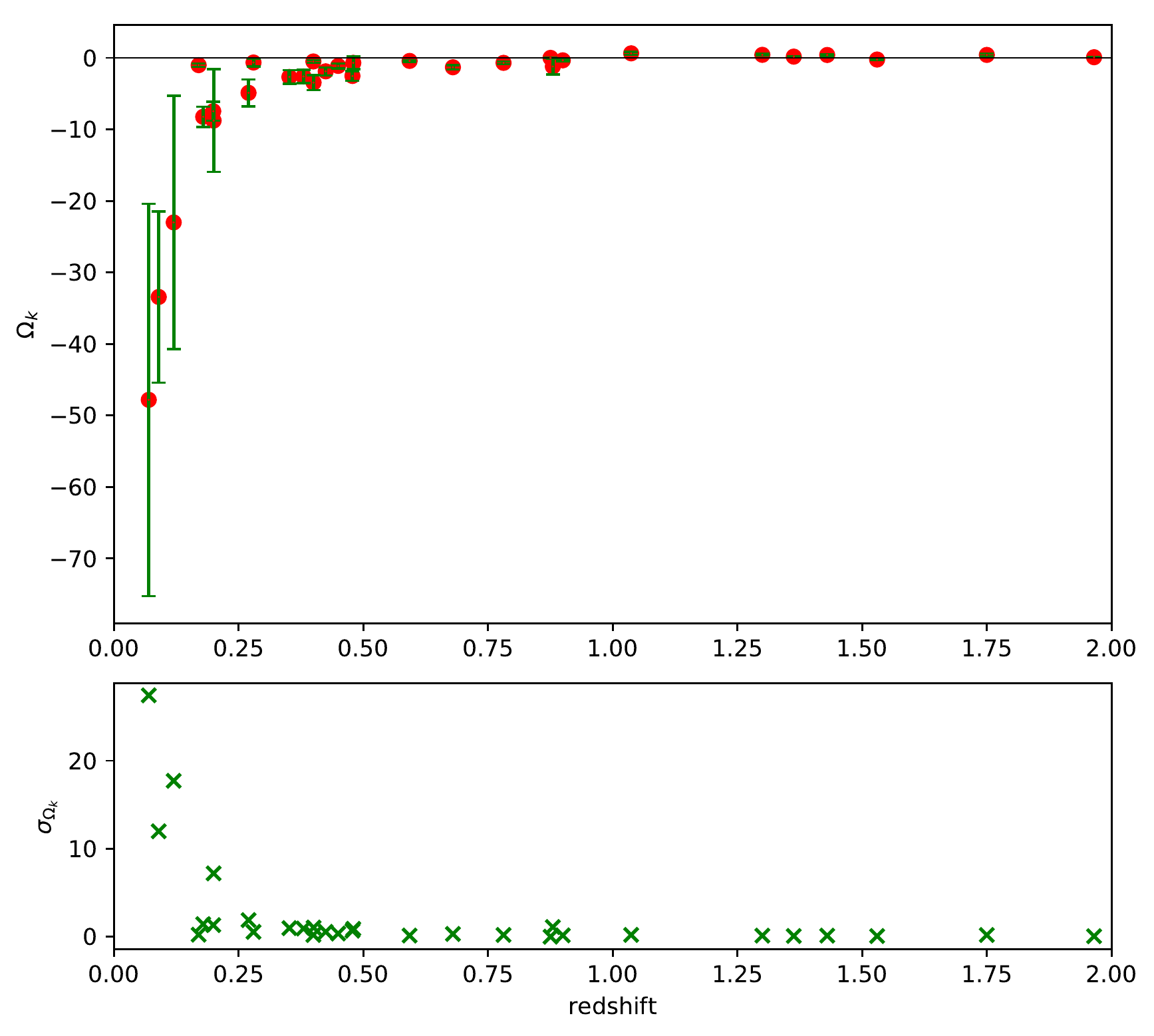}
\caption{Upper panel: $30$ measurements of the spatial curvature parameter $\Omega_k$, shown as red circles 
with $1\sigma$ error bars. Lower panel: the $1\sigma$ errors shown in greater detail.}\label{fig:result}
\end{figure}

The outcome of our calculated spatial curvature parameter $\Omega_k$, based on equation~(\ref{equ:1}),
is shown in figure~\ref{fig:result}. The accuracy of these measurements improves with increasing 
redshift $z$; the level of accuracy is limited by the quality of the data. Not surprisingly, our
results become more consistent with the flatness inferred from the cosmic microwave background
when we sample larger redshifts, spanning distances over which systematic effects from local 
inhomogenieties are smoothed out. At low $z$, we find that $\Omega_k$ is biased towards negative
values, differing from zero by more than $1\sigma$, while at high $z$, the relative error
in $\Omega_k$ is much reduced (see bottom panel in fig.~3). All of the high-$z$ values are
consistent with zero to better than $1\sigma$. 

We tracked the source of error step by step and found that, at low $z$, the uncertainty is
mainly due to errors in the HII-galaxy data. A tiny change in the reconstructed distance modulus 
would result in a very significant change to these nearby $\Omega_k$ results, even allowing
them to flip from positive to negative. There is no such effect at high $z$, however, with
$\Omega_k$ remaining consistent with zero. The low-$z$ determinations of $\Omega_k$ are too
sensistive to errors in the data, and we therefore view these as being not credible compared
to their high-$z$ counterparts. 

We express our results in terms of weighted averages, 
\begin{equation}
\Omega_{k}=\frac{\sum_{i}\Omega_{k,i}/\sigma_{\Omega_{k,i}}^{2}}{\sum_{i}1/\sigma_{\Omega_{k,i}}^{2}},\label{equ:8} 
\end{equation}
where $\Omega_{k,i}$ and $\sigma_{\Omega_{k,i}}$ are the individually determined flatness parameters
and their errors. The corresponding uncertainty in our weighted $\Omega_k$ is estimated in the same way:
\begin{equation}
\sigma_{\Omega_{k}}^{2}=\frac{1}{\sum_{i}1/\sigma_{\Omega_{k,i}}^{2}}\;.\label{equ:9}
\end{equation} 

Reclassifing these data into 4 (redshift) sub-samples, $z > 0$, $z > 0.5$, $z > 1.0$, $z > 1.5$, allows
us to gauge the impact of large-scale smoothing with greater precision. The averaged results are
shown in Table~\ref{table:2}. We see clearly that the ratio $\Omega_{k}/\sigma_{k}$ gets smaller 
with increasing redshift cutoff. The highlights of this work are (1) that $\Omega_{k}=-0.0013\pm0.0004$ 
when all of the currently available data are used, while the most reliable sub-sample, with $z>1.5$,
yields a measurement $\Omega_{k}=-0.01\pm0.04$, fully consistent with a flat Universe, as implied
by the most recent {\it Planck} data \citep{plank1}. We expect that as the precision of the
HII Galaxy data improves, especially at high redshifts, our approach will yield an even
more accurate determination of $\Omega_k$. 

\begin{table}
\begin{center}
\caption{Weighted averages of $\Omega_k$ within various redshift bins}
\renewcommand\tabcolsep{15pt}{
\begin{tabular}{ccc}
\hline
$z$ & $\Omega_{k}$ &$\sigma_{\Omega_{k}}$ \\
\hline
0.0 - 2.0 &-0.001344 &0.000430 \\
0.5 - 2.0 &-0.001330 &0.000430 \\
1.0 - 2.0 & 0.104908 &0.033494 \\
1.5 - 2.0 &-0.011062 &0.041631 \\
\hline
\end{tabular}}\label{table:2}
\end{center}
\end{table}

\section{Conclusion}
We have introduced a new model-independent method of calculating the curvature parameter
$\Omega_k$ using a sample of HII Galaxies and cosmic chronometers, based on the work of \citet{cla1}. 
Since no cosmology is needed to infer $H(z)$ from the latter, the data used in this
approach are unbiased by the choice of model, which in the end might not be completely correct.
In addition, we have inferred the comoving distance as a function of redshift using
a standard candle constructed with HII Galaxies, whose parametrization has also been 
shown to be insensitive to the underlying cosmology. To use this approach, one therefore 
does not need to rely on a fiducial cosmology and assumed priors. 

The flatness parameter is proportional to the balance of energy locally, i.e., the
sum of positive expansion kinetic energy and negative potential energy. Spatial flatness
implies a net zero energy, arguably the most `elegant' initial condition at the big
bang. Such a condition might even be viewed as some supporting evidence in favour
of a quantum fluctuation in vacuum for the origin of the Universe. Our analysis confirms
the flatness inferred by {\it Planck}, thereby possibly obviating---at least with
this level of accuracy---the need to consider new physics associted with an initial
condition of non-zero energy. Our expectation is that the acquisition of a more
accurately measured HII Galaxy sample, particularly at high $z$, should provide an
even more stringent constraint on $\Omega_k$ than we have obtained here. 
\acknowledgments
We are grateful to Cheng-Zong Ruan for
providing a useful advice on reducing error. This work was supported by the National Science Foundation of China (Grants No. 11573006, 11528306), the Fundamental Research Funds for the Central Universities and the Special Program for Applied Research on Super Computation of the NSFC-Guangdong Joint Fund (the second phase). National Key R$\&$D Program of China (2017YFA0402600).


\begin{thebibliography}{}
\bibitem[Blake et al.(2012)]{bla1} Blake, C., Brough, S., Colless, M., et al.\ 2012, \mnras, 425, 405 
\bibitem[Bolton et al.(2008)]{bol} Bolton, A.~S., Burles, S., Koopmans, L.~V.~E., et al.\ 2008, \apj, 682, 964 
\bibitem[Cai et al.(2016)]{cai1}Cai, R.-G., Guo, Z.-K., \& Yang, T., 2016, PRD, 93, 043517
\bibitem[Ch\'avez et al.(2012)]{cha1} Ch\'avez R., Terlevich E., Terlevich R., Plionis M., Bresolin F., Basilakos S., Melnick J., 2012, MNRAS, 425, L56
\bibitem[Ch\'avez et al.(2014)]{cha2} Ch\'avez R., Terlevich R., Terlevich E., Bresolin F., Melnick J., Plionis M., Basilakos S., 2014, MNRAS, 442, 3565
\bibitem[Clarkson et al.(2007)]{cla1}Clarkson, C., Corts, M., \& Bassett, B. 2007, JCAP, 8, 011
\bibitem[Clarkson et al.(2008)]{cla2}Clarkson, C., Bassett, B., \& Lu, T. H.-C. 2008, Physical Review Letters, 101, 011301
\bibitem[Gazta\~nag et al.(2009)]{gaz1}Gazta\~naga, E., Cabr\'e, A., \& Hui, L. 2009, MNRAS, 399, 1663
\bibitem[Jimenez \& Loeb(2002)]{jl2}Jimenez, R., \& Loeb, A., 2002, ApJ, 573, 37
\bibitem[Jimenez et al.(2003)]{jim1}Jimenez, R., Verde, L., Treu, T., Stern, D., 2003. ApJ, 593, 622
\bibitem[Gong \& Wang(2007)]{gw}Gong, Y. G., Wang, A. 2007, PRD, 75, 043520
\bibitem[Hogg(1999)]{hog}Hogg, D. W. 1999, ArXiv Astrophysics e-prints, astro-ph/9905116
\bibitem[Ichikawa et al.(2006)]{ich1}Ichikawa, K., et al. 2006, JCAP, 12, 005
\bibitem[Leaf \& Melia(2017)]{lm}Leaf, K. \& Melia, F. 2017, MNRAS 470, 2320
\bibitem[Li et al.(2014)]{li1}Li, Y.-L., Li, S.-Y., Zhang, T.-J., \& Li, T.-P. 2014, ApJL, 789, L15
\bibitem[Li et al.(2016)]{lzx} Li, Z.-X., Wang, G.-J., Liao, K., \& Zhu, Z.-H.\ 2016, \apj, 833, 240 
\bibitem[Moresco et al.(2012)]{mor1}Moresco M. et al., 2012, JCAP, 8, 006
\bibitem[Moresco(2015)]{mor3}Moresco M., 2015, MNRAS, 450, L16
\bibitem[Moresco et al.(2016a)]{mor2}Moresco M. et al., 2016a, JCAP, 05, 014
\bibitem[M\"ortsell \& J\"onsson(2011)]{mj}M\"ortsell, E., \& J\"onsson, J., 2011, arXiv:1102.4485
\bibitem[Planck Collaboration et al.(2018)]{plank1} Planck Collaboration, Aghanim, N., Akrami, Y., et al.\ 2018, arXiv:1807.06209
\bibitem[Rana et al.(2016)]{ran1}Rana, A., Jain, D., Mahajan, S., Mukherjee, A., 2016, JCAP, 07, 026
\bibitem[Samushia et al.(2013)]{sam1}Samushia, L., Reid, B. A., White, M., et al. 2013, MNRAS, 429, 1514
\bibitem[Sapone et al.(2014)]{sap1}Sapone, D., Majerotto, E., \& Nesseris, S., 2014, PRD, 90, 023012
\bibitem[Seikel et al.(2012)]{sei1}Seikel, M., Clarkson , C., and Smith, M., 2012, JCAP, 06, 036 
\bibitem[Shafieloo \& Clarkson(2010)]{sc}Shafieloo, A. \& Clarkson, C. PRD, 81, 083537
\bibitem[Simon et al.(2005)]{sim1}Simon J., Verde L., Jimenez R., 2005, PRD, 71, 123001
\bibitem[Stern et al.(2010)]{ste1}Stern D., Jimenez R., Verde L., Stanford S. A., Kamionkowski M., 2010,
ApJS, 188, 280
\bibitem[Suzuki et al.(2012)]{suz} Suzuki, N., Rubin, D., Lidman, C., et al.\ 2012, \apj, 746, 85 
\bibitem[Terlevich et al.(2015)]{ter1} Terlevich R.,Terlevich E.,Melnick J.,Ch\'avez R.,Plionis M.,Bresolin F., Basilakos S., 2015, MNRAS, 451, 3001
\bibitem[Virey et al.(2008)]{vir1}Virey, J. M., et al. 2008, JCAP, 12, 008
\bibitem[Wang et al.(2017)]{zzh} Wang, G.-J., Wei, J.-J., Li, Z.-X., Xia, J.-Q., \& Zhu, Z.-H.\ 2017, \apj, 847, 45
\bibitem[Wei et al.(2016)]{wei1} Wei, J.-J., Xu, X.-F. and Melia, F., 2016, MNRAS 463, 1144
\bibitem[Yahya et al.(2014)]{yah1}Yahya, S., Seikel, M., Clarkson, C., Maartens, R., \& Smith, M.\ 2014, \prd, 89, 023503
\bibitem[Yu \& Wang(2016)]{yw}Yu, H. \& Wang, F.-Y., arXiv: 1605.02483
\bibitem[Zhang et al.(2014)]{zhang1}Zhang, C., Zhang, H., Yuan, S., Liu, S., Zhang, T.-J., Sun, Y.-C., 2014. Res. Astron. Astrophys., 14, 1221
\bibitem[Zheng et al.(2016)]{zheng1}Zheng, X., Ding, X., Biesiada, M., Cao, S., \& Zhu, Z.-H.\ 2016, \apj, 825, 17
\end{thebibliography}
\end{document}